\documentstyle[prl,aps,epsfig,floats,twocolumn]{revtex}
\begin{document}
\draft

\title{\hfill OKHEP-00-06\\
Relativistic Coulomb Resummation in QCD}

\author{
Kimball A. Milton\thanks{E-mail: milton@mail.nhn.ou.edu}
 and Igor L. Solovtsov\thanks{On leave of absence from the
Bogoliubov Laboratory of Theoretical Physics, Joint
Institute for Nuclear Research, Dubna, 141980 Russia.
E-mail: solov@mail.nhn.ou.edu, solovtso@thsun1.jinr.ru
}   }
\address{Department of Physics and Astronomy, University of
Oklahoma, Norman, OK 73019 USA}

\date{\today}
\maketitle

\begin{abstract}
A relativistic Coulomb-like resummation factor in QCD is suggested,
based on the solution of the quasipotential equation.
\end{abstract}

In describing a charged particle-antiparticle system near threshold, it
is well known from QED that the so-called Coulomb resummation factor
plays an important role \cite{Schwinger70}. This resummation,
performed on the basis of the
nonrelativistic Schr\"odinger equation with the Coulomb potential
$V(r)=-\alpha/r$, leads to the Sommerfeld-Sakharov
factor~\cite{Sommerfeld,Sakharov}
\begin{equation} \label{SS-factor} 
S_{\rm nr}\,=\,\frac{X_{\rm nr}}{1\,-\,\exp(-X_{\rm nr})}\, , 
\quad\quad X_{\rm nr}\,=\,\frac{\pi\,\alpha}{v_{\rm nr}}\, , 
\end{equation} 
which is related to the wave function of the continuous spectrum at the origin,
$|\psi(0)|^2$. Here $v_{\rm nr}$ is the velocity of the particle.
An expansion of Eq.~(\ref{SS-factor}) in a power series in the
coupling constant $\alpha$ reproduces the threshold singularities of the
Feynman diagrams in the form $(\alpha/v)^n$. However, in 
the threshold region one
cannot truncate the perturbative series and the $S$-factor should be taken
into account in its entirety. A description of quark-antiquark systems near
threshold, which has now been 
intensively investigated \cite{resum}, also requires
this Coulomb resummation. The $S$-factor  appears in the parametrization of
the imaginary part of the quark current correlator, the Drell ratio $R(s)$,
which can be approximated in terms of the Bethe-Salpeter (BS) amplitude 
of two charged particles $\chi_{\rm
BS}(x)$ at $x=0$~\cite{BarbieriCR73}. The nonrelativistic replacement of
this amplitude by the wave function which obeys the Schr\"odinger equation
with the Coulomb potential, leads to the approximation~(\ref{SS-factor}) with 
$\alpha\to 4\alpha_s/3$,  for QCD.  

In the relativistic theory, especially for systems composed of quarks lighter 
than the top, 
the nonrelativistic approximation needs to be modified.
To use the $S$-factor within such a relativistic regime one usually uses the
simple substitution $v_{\rm nr}\to v$ with $v=\sqrt{1-{4m^2}/{s}}$. However, the
corresponding relativistic generalization of the $S$-factor is obviously
not unique, for there are numerous ways of expressing the nonrelativistic
velocity in terms of the relativistic energy $\sqrt{s}$. For a systematic
relativistic analysis of  quark-antiquark systems, it is essential
from the very beginning to have a relativistic generalization of the $S$-factor.

In this letter we suggest a new form for this
relativistic factor in the case of QCD. Our
starting point is the quasipotential (QP) approach proposed by Logunov and
Tavkhelidze \cite{LogunovTav}, in the form
suggested by Kadyshevsky~\cite{Kadyshevsky68}. To find an explicit form
for the relativistic $S$-factor we will transform the QP equation from
momentum space into relativistic configuration
space~\cite{KadMS68-72}. The local Coulomb potential defined in this
representation has a QCD-like behavior in momentum
space~\cite{SavrinS80}.

The possibility of using the QP approach to define the relativistic $S$-factor
is based on the fact that the BS amplitude, which parameterizes the
physical quantity $R(s)$, is taken at $x=0$, therefore, in
particular, at relative time $\tau=0$. The QP wave function is
defined as the BS amplitude at $\tau=0$, and the $R$-ratio can be
expressed through the QP wave function $\psi_{\rm QP}({\bf p})$ by using the
relation
\begin{equation}
\label{BS-QP}
\chi_{\rm BS}(x=0)\,=\,\int d\Omega_p \,\psi_{\rm QP}({\bf p})\, ,
\end{equation}
where $d\Omega_p=(d{\bf p})/[(2\pi)^3\,E_p]$ 
is the relativistic three-dimensional volume element in 
the Lobachevsky space realized on the hyperboloid $E_p^2-{\bf p}^2=m^2$.%
\footnote{In the following we will consider the case of two scalar particles
with the same masses $m$ and use the system of units $c=\hbar=m=1$.}

The QP equation in momentum space has the form
\begin{equation}
\label{OP-m-eq}
\left(2E \,-\,2E_p\right)\,\psi ({\bf p})\,=\,
\int\,d\Omega_k\,V\left({\bf p}(-){\bf k}\right)\,\psi ({\bf k})\, .
\end{equation}
The proper Lorentz transformation, $\Lambda_{\bf k}$, means
a translation in the Lobachevsky space
\begin{equation}
\label{translation}
\Lambda_{\bf k}\,{\bf p}\,\equiv\,{\bf p}(+){\bf k}
\,=\,{\bf p}+{\bf k}\left[ \sqrt{1+{\bf p}^2}
\,+\,\frac{{\bf p}\cdot{\bf k}}{1+\sqrt{1+{\bf k}^2}}\right]\, .
\end{equation}
The role of the plane waves corresponding to the 
translations (\ref{translation}) is played by the following functions
\begin{equation}
\label{xi-funct}
\xi({\bf p},{\bf r})\,=\,\left(E_p-{\bf p}\cdot{\bf n}\right)^{-1-ir}\, ,
\end{equation}
where ${\bf r}={\bf n}r$ and ${\bf n}^2=1$. These functions correspond to the
principal series of unitary representations of the Lorentz group and in the
nonrelativistic limit ($p\ll1$, $r\gg1$)
$\xi({\bf p},{\bf r})\to\exp(i{\bf p}\cdot{\bf r})$.
The orthogonality and completeness relations for these functions are
\begin{eqnarray}
\label{orthogonality}
\int\,d\Omega_p\, \xi({\bf p},{\bf r})\,\xi^*({\bf p},{\bf r}')
&=&\delta({\bf r}-{\bf r}')\,, \nonumber\\
\int (d{\bf r})\, \xi({\bf p},{\bf r})\,\xi^*({\bf k},{\bf r})
&=&(2\pi)^3\delta({\bf p}(-){\bf k})   \, ,
\end{eqnarray}
where the relativistic momentum-space $\delta$-function is
$\delta({\bf p}(-){\bf k})=\sqrt{1+{\bf p}^2}\,\delta({\bf p}-{\bf k})$.
The QP wave functions in the momentum and
relativistic configuration representations are related as follows:
\begin{eqnarray}
\label{p-r-funct}
\psi ({\bf r})&=&
\int d\Omega_p\,\xi({\bf p},{\bf r})\,\psi ({\bf p})\,,\nonumber\\
\psi ({\bf p})&=&
\int (d{\bf r})\,\xi^*({\bf p},{\bf r})\,\psi ({\bf r})\,.
\end{eqnarray}

For a spherically symmetric potential the $\xi$-transform of
Eq.~(\ref{OP-m-eq}) is the equation
\begin{eqnarray}
\label{r-eq1}
&&\int\,d\Omega_p\, (d{\bf r}')\,(2E -2E_p)\, \xi({\bf p},{\bf r})
\xi^*({\bf p},{\bf r}')\,\psi ({\bf r}')\nonumber\\
&&\quad=V(r)\,\psi ({\bf r}),
\end{eqnarray}
where the  right hand side is local.  Here the transform of the potential
is given in terms of the same relativistic plane wave,
\begin{equation}
V({\bf p(-)k})=\int(d{\bf r})\,\xi^*({\bf p(-)k,r})V({\bf r}).
\end{equation}
The left hand side of this equation can be rewritten in a non-integral form
by using the operator of the free Hamiltonian \cite{KadMS68-72}
\begin{equation}
\label{H-0}
\hat{H}_0\,=\,\cosh\left(i\frac{d}{dr}\right) 
+\frac{i}{r}\sinh\left(i\frac{d}{dr}\right) 
-\frac{\Delta_{\theta,\varphi}}{2r^2}\exp\left(i\frac{d}{dr}\right)\,, 
\end{equation}
where $\Delta_{\theta,\varphi}$ is the angular part of the Laplacian
operator.
The relation
$\hat{H}_0\xi({\bf p},{\bf r})=E_p\xi({\bf p},{\bf r})$
allows one to re-express the equation in terms of finite differences
\begin{equation}
\label{r-eq-fd}
\left(2E -2\hat{H}_0\right)\psi ({\bf r})\,=\,V(r)\,\psi ({\bf r})\,.
\end{equation}
This equation for the Coulomb potential has been investigated in
Ref.~\cite{Freeman69}. The solutions contain arbitrary
functions of $r$ with period $i$, 
the so-called the $i$-periodic constants, which
appear in the solutions due to the finite difference nature of the Hamiltonian
(\ref{H-0}). For some problems, such as defining the bound
state spectrum, this $i$-periodic constant is not important. However, for
the purpose of extracting the $S$-factor,
we must develop a method which avoids this ambiguity.

Consider the Coulomb potential defined in relativistic configuration
space
\begin{equation}
\label{Coulomb}
V(r)\,=\,-\frac{\alpha}{r}\,.
\end{equation}
The $\xi$-transformation of Eq.~(\ref{Coulomb}) gives in momentum space 
the following function
\begin{equation}
\label{Coulomb-moment}
V(\Delta)\,\sim\, \frac{1}{\chi_\Delta\sinh\chi_\Delta}\,,
\end{equation}
where the relative rapidity $\chi_\Delta$ corresponds to 
${\bf \Delta}={\bf p}(-){\bf k}$
and is defined in terms of the square of the momentum transfer 
by $Q^2=-(p-k)^2=2(\cosh\chi_\Delta -1)$. 
For large $Q^2$ the potential $V(\Delta)$ behaves as $(Q^2\ln Q^2)^{-1}$,
which reproduces the principal behavior of the QCD potential proportional to
$\bar\alpha_s(Q^2)/Q^2$ with $\bar\alpha_s(Q^2)$ being the QCD running coupling.

According to Eqs.~(\ref{BS-QP}), (\ref{xi-funct}), and
(\ref{p-r-funct}), we find a relation between the required BS amplitude
and the QP wave function,
$\chi_{\rm{BS}}(x=0)=\psi_{\rm{QP}}(r=i)$. Performing a partial-wave analysis
we further observe that the QP wave function for an $\ell$-state will 
contain the generalized power $(-r)^{(\ell+1)}=
i^{l+1}\Gamma(ir+l+1)/\Gamma(ir)$,
 which vanishes at $r=i$ for $\ell\not=0$. Thus, we need only to consider
the $\ell=0$ wave function for which we
can write $\psi ({\bf r})=\psi (r)$.
Introducing the function $R (r)=r\,\psi (r)$ into Eq.~(\ref{r-eq1}), 
we get
\begin{eqnarray}
\label{r-eq2}
&&\frac{2}{\pi}\int_0^\infty d\chi'\int_0^\infty dr'\sin \chi' r\,
\sin \chi'r' (2E -2\cosh\chi') R (r')\nonumber\\
&&\quad=V(r)\, R (r)\,.
\end{eqnarray}

We will seek a solution of Eq.~(\ref{r-eq2}) with the Coulomb
potential (\ref{Coulomb}) in the form 
\begin{equation}
\label{form-of-solut}
R (r)\,=\,\int_\alpha^\beta\,d\zeta\,\exp(i\,r\,\zeta)\,R (\zeta)\,,
\end{equation}
where the $\zeta$-integration is performed in the complex plane over a
contour with endpoints $(\alpha,\beta)$ \cite{SkSol83}. Substituting
Eq.~(\ref{form-of-solut}) into Eq.~(\ref{r-eq2}) we find the
equation\footnote{We first perform the $\chi'$-integration with a
regularization factor $\exp(-\epsilon\chi^{\prime2})$ 
and then set $\epsilon=0$ after
all calculations.}
\begin{eqnarray}
\label{r-eq4}
&&\int_\alpha^\beta\,d\zeta\,\exp(i\,r\,\zeta)\,
\left(\,2E -2\cosh\zeta\right)\,R (\zeta)\nonumber\\
&&\quad=-\frac{\alpha}{r}\,
\int_\alpha^\beta\,d\zeta\,\exp(i\,r\,\zeta)\, R (\zeta)\,,
\end{eqnarray}
which, when we integrate by parts, yields the two equations
\begin{equation}
\label{nonintegral-term}
\exp(i\,r\,\zeta)\,\left(\,2E -2\cosh\zeta\right)
\,R (\zeta)\,\Big|_{\zeta=\alpha}^{\zeta=\beta}\,=\,0 
\end{equation}
and
\begin{equation}
\label{integral-term}
i\,\frac{d}{d\zeta}\,
\Big[\left(\,2E -2\cosh\zeta\right)\,R (\zeta)\Big]\, 
=\,-\,\alpha\,R (\zeta) \,.
\end{equation}
The solution of Eq.~(\ref{integral-term}) is
\begin{eqnarray}
\label{R-zeta}
R (\zeta)&=&C(\chi)\, \exp(\zeta)
\Big[\exp(\zeta)\,-\,\exp(-\chi)\Big]^{A-1}\nonumber\\
&&\quad\times\Big[\exp(\zeta)\,-\,\exp(\chi)\Big]^{-A-1} \,,
\end{eqnarray}
where $A=i\alpha/(2\sinh\chi)$, $E =\cosh\chi$, and $C(\chi)$ is an arbitrary
function of $\chi$.

	  \begin{figure}
\centerline{ \epsfig{file=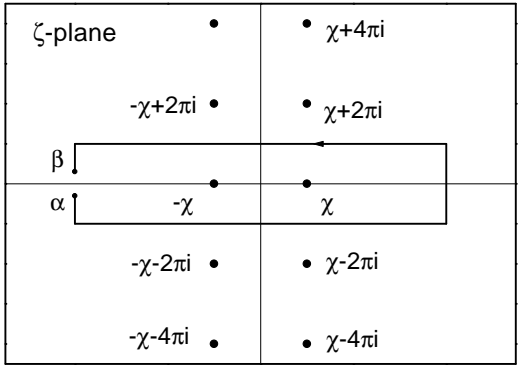,width=3.5in}}
\caption{\sl Contour of integration in Eq.~(\protect\ref{form-of-solut})
and singularities of the function (\protect\ref{R-zeta})
in the complex $\zeta$-plane. }
\label{zeta-plane}
\end{figure}

The branch points of the function (\ref{R-zeta}) are $\pm\chi+2\pi in$ (see
Fig.~\ref{zeta-plane}). The contour of integration must not intersect cuts
which we take from $-\infty+2\pi i n$ to $\pm\chi+2\pi i n$. In the case
when the interaction vanishes, $\alpha\to0$, the solution $R (r)$ should
reproduce the known 
free wave function ${\sin\chi r}/{\sinh\chi}$. Taking into account
these remarks and Eq.~(\ref{nonintegral-term}) for the boundary values at
$(\alpha,\beta)$, we take $\alpha=-R-i\varepsilon$, $\beta=-R+i\varepsilon$
with $R\to\infty$. The vertical part of the contour to the right is given
by ${\rm Re}\,\zeta=+R$. It
is also convenient for finding a connection to an integral representation
of the hypergeometric function to take the horizontal parts of the contour
to be characterized by ${\rm Im}\,\zeta=\pm\pi$ (see Fig.~\ref{zeta-plane}).

The resulting solution does not contain the $i$-periodic constant and reads%
\footnote{ The representation of this solution in terms of the hypergeometric
function can be found, for instance, by the substitution $x=\chi-\ln s$.}
\begin{eqnarray}
\label{R(r)1}
&&R (r)=C(\chi)\,\sinh\pi r\,\int_{-\infty}^\infty\,
dx\,\exp(irx)\,\exp(x)\nonumber\\
&&\times\Big[\exp(x)\,+\,\exp(-\chi)\Big]^{A-1}
\Big[\exp(x)\,+\,\exp(\chi)\Big]^{-A-1}.\nonumber\\
\end{eqnarray}
Comparing the asymptotic form of Eq.~(\ref{R(r)1}) at $r\to\infty$ with the
free wave function we can determine the constant $C(\chi)$ and 
calculate $|\psi_{\rm QP}(i)|^2$ which leads to the relativistic $S$-factor:
\begin{equation}
\label{S-factor-relativistic}
S(\chi)\,=\,\frac{X(\chi)}{1\,-\,\exp\left[-X(\chi)\right]}\, ,
\quad\quad X(\chi)\,=\, \frac{\pi\,\alpha}{\sinh\chi}\, ,
\end{equation}
where $\chi$ is the rapidity which related to $s$ by
$2\cosh\chi=\sqrt{s}$. The function $X(\chi)$ in
Eq.~(\ref{S-factor-relativistic})  can be expressed in terms of
$v$ as  $X(\chi)=\pi\alpha\sqrt{1-v^2}/v$.

\begin{figure}
\centerline{ \epsfig{file=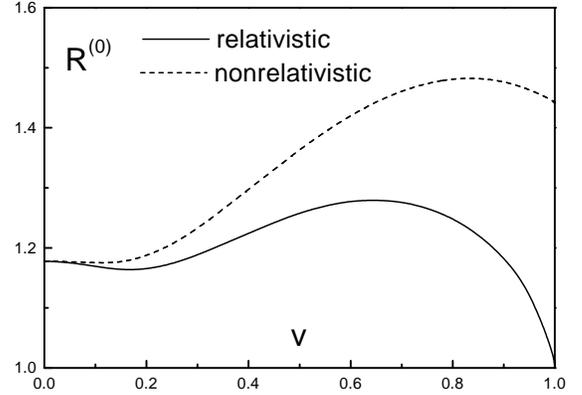,width=3.5in}}
\caption{\sl Behavior of $R^{(0)}(s)$ with relativistic and nonrelativistic
$S$-factors.}
\label{fig2}
\end{figure}

We note that this new relativistic factor
could have a significant impact in interpreting strong-interaction
physics.  In many physically interesting cases, $R(s)$ occurs as a factor
in an integrand, as, for example, for the case of inclusive $\tau$ decay,
for smearing quantities, and for the Adler $D$ function.  Here the behavior
of $S$ at intermediate values of $v$ becomes important.  
To illustrate the difference between the factors~(\ref{SS-factor})
and (\ref{S-factor-relativistic}), in Fig.~\ref{fig2}
we plot the principal contribution to $R(s)$ for the vector currents,
\begin{equation}
R^{(0)}(s)={v(3-v^2)\over2}S,
\end{equation}
for the nonrelativistic case with $v_{\rm nr}\to v$ and the relativistic
one, for $\alpha=0.25$.

We conclude this letter by discussing the two limiting cases.
In the nonrelativistic limit, $v\ll1$, the relativistic $S$-factor
(\ref{S-factor-relativistic}) reproduces the nonrelativistic result
(\ref{SS-factor}). In the ultrarelativistic limit, as it has been argued in
Ref.~\cite{Lucha}, the bound state spectrum vanishes as $m\to 0$ because the
particle mass is the only dimensional parameter. This feature reflects an
essential difference between  potential models and quantum field theory,
where an additional dimensional parameter appears. One can conclude that
within a potential model, the $S$-factor which corresponds to the continuous 
spectrum should go to unity in the limit $m\to 0$. Thus, the
relativistic resummation factor $S$ obtained here reproduces both the expected
nonrelativistic  and ultrarelativistic limits and
corresponds to a QCD-like Coulomb potential.

\section*{Acknowledgments}

The authors would like to thank A.N.~Sissakian, D.V.~Shirkov, and
O.P.~Solovtsova for interest in this work and valuable discussions. Partial
support of the work by the US National Science Foundation, grant PHY-9600421,
by the US Department of Energy, grant DE-FG-03-98ER41066, and by the RFBR,
grants 99-01-00091, 99-02-17727, is gratefully acknowledged. The work of ILS
was also supported in part by the University of Oklahoma, through its College
of Arts and Science, the Vice President for Research, and the Department of
Physics and Astronomy. He thanks the 
members of the high energy group of the University of
Oklahoma for their warm hospitality.


\end{document}